\documentclass[conference]{IEEEtran}
\IEEEoverridecommandlockouts
\usepackage{cite}
\usepackage{amsmath,amssymb,amsfonts}
\usepackage{graphicx}
\usepackage{textcomp}
\usepackage{xcolor}
\usepackage[letterpaper, left=0.625in, right=0.625in, bottom=1in, top=0.75in]{geometry}
\usepackage[hyperfootnotes=false,hidelinks]{hyperref}
\usepackage[subrefformat=parens,labelformat=parens,caption=false]{subfig}
\usepackage[utf8]{inputenc}
\usepackage{graphicx}
\usepackage{cite}
\usepackage{amsmath,amssymb,amsfonts}
\usepackage{psfrag}
\usepackage{graphicx}
\usepackage{textcomp}
\usepackage{xcolor}
\usepackage{lipsum}
\usepackage{mathtools}
\usepackage{cuted}
\usepackage{stfloats}
\usepackage{siunitx}
\usepackage{algorithm}
\usepackage{algorithmic}
\usepackage{amsmath}
\usepackage{amsfonts}
\usepackage{amssymb}
\usepackage{balance}
\usepackage{scalerel}
\usepackage{tikz}
\usetikzlibrary{svg.path}
\usepackage{graphicx} 
\usepackage{booktabs} 
\usepackage{wasysym}
\usepackage{amsmath,amsfonts}
\usepackage[scr=boondox]{mathalpha}

\newcommand{\Ccal}{\mathcal{C}}

\newcommand{\Ncal}{\mathcal{N}}

\newcommand{\Qcal}{\mathcal{Q}}

\newcommand{\Rbb}{\mathbb{R}}

\newcommand{\Nbb}{\mathbb{N}}

\DeclareMathAlphabet\mathbfcal{OMS}{cmsy}{b}{n}

\renewcommand{\IEEEQED}{\IEEEQEDopen}

\def\BibTeX{{\rm B\kern-.05em{\sc i\kern-.025em b}\kern-.08em
    T\kern-.1667em\lower.7ex\hbox{E}\kern-.125emX}}
\begin{document}

\title{Distributed Maximum Consensus over Noisy Links}

\author{
\IEEEauthorblockN{
{Ehsan Lari}\textsuperscript{1},
\text{Reza Arablouei}\textsuperscript{2},
\text{Naveen K. D. Venkategowda}\textsuperscript{3},
\text{Stefan Werner}\textsuperscript{1}
}
\textsuperscript{1}\text{Department of Electronic Systems, Norwegian University of Science and Technology, Trondheim, Norway} \\
\textsuperscript{2}\text{CSIRO’s Data61, Pullenvale QLD 4069, Australia}\\
\textsuperscript{3}\text{Department of Science and Technology, Linköping University, Norrköping, Sweden} \\
\thanks{This work was supported by the Research Council of Norway.} 
}

\maketitle

\begin{abstract}

We introduce a distributed algorithm, termed noise-robust distributed maximum consensus (RD-MC), for estimating the maximum value within a multi-agent network in the presence of noisy communication links. Our approach entails redefining the maximum consensus problem as a distributed optimization problem, allowing a solution using the alternating direction method of multipliers. Unlike existing algorithms that rely on multiple sets of noise-corrupted estimates, RD-MC employs a single set, enhancing both robustness and efficiency. To further mitigate the effects of link noise and improve robustness, we apply moving averaging to the local estimates. Through extensive simulations, we demonstrate that RD-MC is significantly more robust to communication link noise compared to existing maximum-consensus algorithms.

\end{abstract}

\section{Introduction} \label{intro}

Distributed learning algorithms have garnered significant attention in recent years for addressing data-centric challenges across large-scale multi-agent networks. These algorithms find diverse applications across various analytics tasks~\cite{1605401,5422758,6573232,4787093,6805223,4407653,6805127}. Distributed algorithms not only enjoy enhanced resilience against node or link failures, compared to centralized algorithms, but also obviate the need for central data collection and processing.

Consensus algorithms play a pivotal role in a variety of distributed computing and optimization applications, including those related to distributed learning~\cite{ruan2019secure,9795093,9954449,10068288,10097514}. These algorithms facilitate coordination and consensus formation among multiple agents within a distributed system, enabling them to collaboratively achieve a common goal. Therefore, they serve as a fundamental component in systems reliant on distributed decision-making~\cite{4118472,7902114,10032214}. Several studies~\cite{7054482,7398012,9825718,10383290,9881550,9940480,montijano2014robust,7511707,9683179} delve into the problem of attaining network-wide consensus on various values such as average, minimum, and median in a distributed manner. Achieving consensus in a multi-agent network mandates local computations by agents, coupled with data exchange among neighboring agents. Hence, the presence of communication noise, especially over wireless links, necessitates careful consideration. 

The distributed maximum consensus problem pertains to identifying the maximum value within a network. Extensive research has been conducted on this problem in various contexts~\cite{8264362,naveen5,7574375,8859263,9217911,9816009,9650547}. For instance, \cite{8264362} presents a distributed algorithm for maximum consensus, albeit assuming noiseless links. In addition,~\cite{naveen5} derives bounds on the expected convergence time for maximum consensus in asynchronous networks without considering communication noise. The approach in~\cite{7574375} addresses the maximum consensus problem by approximating the maximum function with the soft-max function. However, its performance is limited by a trade-off between estimation error and convergence speed. While \cite{8859263} proposes a noise-robust distributed maximum consensus algorithm, its error variance increases linearly with the network size.

In this paper, we introduce a fully-distributed algorithm, called noise-robust distributed maximum consensus (RD-MC), devised to accurately estimate the maximum value across a multi-agent network, particularly in scenarios where communication channels are corrupted by noise. In developing RD-MC, we draw inspiration from previous research~\cite{4407653,sspehsan,apsipaehsan} that highlight the benefits of strategically designed parameter exchanges. We substantiate the effectiveness of RD-MC by conducting extensive simulations and comparing its performance with existing algorithms.

\textit{Mathematical Notations:} The sets of natural and real numbers are denoted by $\Nbb$ and $\Rbb$, respectively. Scalars and column vectors are denoted by lowercase and bold lowercase, respectively. The indicator function $\mathcal{I}_{a}(x)$ is defined as $ \mathcal{I}_{a} (x) = 0 $, if $x \geq a$ and $\infty$ otherwise.

\section{Preliminaries} \label{preli}

We consider a connected network comprising $J\in\Nbb $ agents and $E\in\Nbb $ edges, modeled by an undirected graph $\mathcal{G(V,E)}$. Here, the set of vertices $\mathcal{V}=\{1,2,\cdots,J\}$ corresponds to the agents and the edge set $\mathcal{E}$ represents the communication links between the agents. Agent $i\in\mathcal{V}$ communicates with its neighbors, indexed in $\mathcal{N}_{i}$ with cardinality $d_i=|\mathcal{N}_{i}|$. The set $\mathcal{N}_{i}$ does not include the agent $i$ itself. We consider only simple graphs, devoid of self-loops or multiple edges. The structure of $\mathcal{G}$ is described by its adjacency matrix $\mathbf{A}$ with entries $\mathscr{a}_{ij}$, where $\mathscr{a}_{ij} = 1 $ if $ (i,j) \in \mathcal{E} $ and $\mathscr{a}_{ij} = 0 $ if $ (i,j) \notin \mathcal{E} $. Furthermore, the degree matrix $\mathbf{D}=\mathrm{diag}(d_1,\cdots,d_J)$ contains the number of nodes in each agent's neighborhood.

The conventional maximum consensus algorithm, which relies on the network agents communicating solely with their immediate neighbors, is expressed as
\begin{equation}
    x_{i}({k+1}) = \max (x_{i}(k),\{x_{j}(k)\}_{j \in \mathcal{N}_{i}}), \quad \quad \forall i \in \mathcal{V},
    \label{eq:maxcons1}
\end{equation}
where $ x_{i}({k}) $ is the estimate of the $i$th agent at time instant $k$, and the initial value is $x_{i}(0) = a_{i}\ \forall i\in\mathcal{V}$. We denote the solution to \eqref{eq:maxcons1} by $\max(\{a_{i}\}_{i\in\mathcal{V}}) = a^{\star}$. When the inter-node communications are noiseless, there exists a finite $K$ such that $x_{i}(k) = a^{\star}\ \forall k\geq K$ and $i\in\mathcal{V}$~\cite{9217911}.

We consider the communication links to be noisy. We denote the noise in the message received by agent $i$ from agent $j$ at time instant $k$ as ${w}_{j}^i({k})\in\Rbb$ and model it as zero-mean additive white Gaussian noise with variance $\sigma^2$. We assume that the noise is uncorrelated across different time instants and agents.
By accounting for additive link noise, \eqref{eq:maxcons1} becomes
\begin{equation}
    x_{i}({k+1}) = \max (x_{i}(k),\{x_{j}(k) +{w}_{j}^i({k})\}_{j \in \mathcal{N}_{i}}),\quad \quad \forall i \in \mathcal{V}.
    \label{eq:convmaxcons}
\end{equation}

While~\eqref{eq:maxcons1} converges to the maximum value under ideal communication conditions, \eqref{eq:convmaxcons} may fail to converge due to potential noise-induced drift in the estimated maximum value during each iteration. This drift, compounded over time, can lead to significant inaccuracies. To address this challenge, we reformulate the maximum consensus problem as a distributed optimization problem with an aggregate global objective function. The proposed RD-MC algorithm, developed to solve this problem, converges even in the presence of additive noise in the communication links.

\section{Noise-Robust Maximum Consensus Algorithm} \label{algorithm}

In this section, we present a reformulation of the maximum consensus problem that enables its solution via ADMM. Subsequently, we describe two subtle modifications to the algorithm resulting from solving the reformulated problem through ADMM. These modifications are aimed at enhancing the robustness of distributed maximum consensus to communication noise and lead to the proposed RD-MC algorithm.

The consensus-based reformulation of the maximum consensus problem~\eqref{eq:maxcons1} has been discussed in~\cite{9217911}. However, the effect of noisy links has not been investigated in that work. In \cite{9217911}, it is demonstrated that \eqref{eq:maxcons1} can be equivalently expressed as
\begin{align} \label{eq9}
 \min_{\{x_{i},y_{i},q_{i}^{j}\}} \quad & \frac{1}{J} \sum_{i=1}^{J} x_{i} + \frac{1}{J} \sum_{i=1}^{J} \mathcal{I}_{a_i} (y_i) \\
\textrm{s.t.} \quad & x_{i} = y_{i} \quad \forall i \in \mathcal{V} \notag \\
& x_{i} = q_{i}^j, x_{j} = q_{i}^j \quad \forall i\in\mathcal{V}, j\in\mathcal{N}_i, \notag 
\end{align}
where the indicator function $\mathcal{I}_{a_i} (y_i)$ imposes an inequality constraint to seek the maximum value and the auxiliary variables $\Qcal = \{q_{i}^j\}_{i\in\mathcal{V},j\in\mathcal{N}_i}$ facilitate consensus within each agent's neighborhood and, consequently, across the network. The optimization problem~\eqref{eq9} can be tackled using various methods, including those based on subgradients or ADMM. However, distributed subgradient methods applied to affine objective functions are known to converge slowly~\cite{4749425}. Therefore, we opt for ADMM to solve \eqref{eq9}.

Let $\mathcal{L}_{\rho}(\{x_{i},y_{i}\}_{i\in\mathcal{V}},\Qcal,\mathcal{M})$ denote the augmented Lagrangian function associated with \eqref{eq9}, where $\mathcal{M} = \{u_i,\mu_i^j,\pi_i^j\}_{i\in\mathcal{V},{j\in\mathcal{N}_i}}$ represents the respective Lagrange multipliers. Minimizing $\mathcal{L}_{\rho}$, while applying the Karush-Kuhn-Tucker optimality conditions~\cite{boyd2004convex} to \eqref{eq9} and defining $v_i(k) = 2 \sum_{j \in \mathcal{N}_{i}}\mu_i^j(k)$, leads to the following iterative updates at the $i$th agent along with the elimination of $\{\pi_i^j\}_{i\in\mathcal{V},{j\in\mathcal{N}_i}}$ and $\Qcal$ \cite{9217911,giannakis2017decentralized}:
\begin{subequations} \label{eq:FDUp}
\begin{align}
	x_i({k+1})& = {n}_i \big(-J^{-1}+\rho_y[y_i({k})-\bar{u}_i({k})] - v_i(k) \notag\\
 & \hspace{4mm} +\rho_z \sum_{j \in \mathcal{N}_{i}} \left[x_i(k) + \tilde{x}_j(k)\right] \big),
\label{eq:FDxUp}
\\ y_{i}({k+1}) &= \mathrm{max}(x_i({k+1})+\bar{u}_i(k),a_i),
\label{eq:FDyUp}
\\  \bar{u}_i({k+1}) &= \bar{u}_i(k) + x_i({k+1}) - y_i({k+1} ),
\label{eq:FDuUp}
\\ v_i({k+1})& = v_i(k) + \rho_z \sum_{j \in \mathcal{N}_{i}} [x_i(k+1) - \tilde{x}_j(k+1)].
\label{eq:FDvUp}
\end{align}
\end{subequations}
Here, $k$ is the iteration index, $n_i=(\rho_y+2\rho_z d_i)^{-1}$, $\rho_y > 0 $ and $\rho_z > 0$ are the penalty parameters, and $\bar{u}_i={u_i}/{\rho_y}$. In addition, all initial values $\{x_i(0),y_i(0),\bar{u}_i(0),v_i(0)\}_{i\in\mathcal{V}}$ are set to zero.
Note that, in \eqref{eq:FDxUp}, agent $i$ has access to $\tilde{x}_j(k)=x_j(k)+w_i^j(k)$ rather than $x_j(k)$. The iterations~\eqref{eq:FDUp} can be implemented locally at each agent in a fully distributed fashion, as the required information is available within each agent’s neighborhood. We refer to this algorithm, originally proposed in~\cite{9217911}, as distributed maximum consensus (D-MC).

Using the initial values $v_i(0)=0\ \forall i\in\mathcal{V}$, we obtain
\begin{equation}\label{v_i}
    v_i(k)=\rho_z \sum_{\ell = 1}^{k} \sum_{j \in \mathcal{N}_{i}} [x_i(\ell) - \Tilde{x}_j(\ell)]
\end{equation}
from~\eqref{eq:FDvUp}. Substituting~\eqref{v_i} into~\eqref{eq:FDxUp} while using the initial values $x_i(0) = 0$ and $x_i(1) = -J^{-1} n_i\ \forall i\in\mathcal{V}$, we can eliminate $v_i(k)$ and modify \eqref{eq:FDxUp} as
\begin{subequations} \label{eq:FDRobust}
\begin{align}
  x_i({k+1}) & = (1 - \rho_y {n}_i) x_i({k}) - \rho_z d_i {n}_i x_i({k-1}) \notag
  \\& \hspace{4mm} + {n}_i \Big[ \rho_y  z_i(k) + \rho_z \sum_{j \in \mathcal{N}_{i}} \tilde{s}_{j}(k)  \Big],
 \label{eq:FDxUp3}
 \\ \bar{x}_{i}({k+1}) & = \sum_{\ell = 0}^{\Ccal-1} \alpha_{\ell} x_i(k+1-\ell),
 \label{eq:FDxBarUp}
  \\ z_i(k+1) & = 2 y_i({k+1}) - y_i({k}),
  \label{eq:FDs}
 \\ {s}_{i}(k+1) & = 2 \bar{x}_{i}({k+1}) - x_i(k).
 \label{eq:FDz}
\end{align}
\end{subequations}
Note that, in \eqref{eq:FDxBarUp}, to enhance robustness against spurious noise, we compute the convex combination of $\Ccal$ past local estimates, utilizing the weights $\alpha_{\ell}$ that sum to one.

In this alternative formulation, instead of ${x}_{i}(k)$, agents exchange ${s}_{i}(k)$, which is a smoothed version of ${x}_{i}(k)$. However, due to communication noise, they receive noisy versions from their neighbors, i.e., agents $j\in\mathcal{N}_i$ receive $\tilde{s}_{i}(k) = {s}_{i}(k) + w_i^j(k)$ from agent $i$. 
The recursions \eqref{eq:FDRobust} alongside \eqref{eq:FDyUp} and \eqref{eq:FDuUp} constitute the proposed noise-robust distributed maximum consensus (RD-MC) algorithm, summarized in Algorithm \ref{Alg1}. 

We mitigate the effect of noisy links in RD-MC through two key modifications to D-MC. First, the introduction of ${s}_{i}(k)$, a linear combination of $\bar{x}_{i}(k)$ and ${x}_{i}(k-1)$, offers a strategic advantage in alleviating the adverse effects of communication noise. By exchanging ${s}_{i}(k)$ instead of ${x}_{i}(k)$ over noisy links, we enhance robustness. Notably, while the aggregation of two sets of noisy estimates received from neighbors at consecutive iterations [i.e., $\tilde{x}_j(k)$ in~\eqref{eq:FDxUp} and $\tilde{x}_j(k+1)$ in~\eqref{eq:FDvUp}] renders D-MC vulnerable to noise accumulation, RD-MC's reliance on a single set of noisy estimates [i.e., $\tilde{s}_j(k)$ in~\eqref{eq:FDxUp3}] enhances its resilience to link noise. 
Second, we further enhance robustness to link noise by applying a weighted averaging of $x_i({k+1})$ over a sliding window of size $\Ccal$ as in \eqref{eq:FDxBarUp}.

\begin{algorithm}[t!]
\textbf{Parameters}: penalty parameters $\rho_{z}$ and $\rho_{y}$\\[1mm]
\textbf{Initialization}: $x_i(0)=0$, $x_i(1) = -J^{-1} n_i$, $\bar{u}_i(1) = 0$,\\
\hspace*{20.6mm} $z_i(1) = 0$, $s_i(1) = -2J^{-1} n_i$, $\forall i\in\mathcal{V}$\\[1mm]
\textbf{For} \,$k=1,\cdots,K$ \textit{until convergence} \\[1mm]
\hspace*{2.5mm} Receive $\tilde{s}_{j}(k)$ from neighbors $j \in \mathcal{N}_{i}$ \\[1mm]
\hspace*{2.5mm} $x_i({k+1}) = (1 - \rho_y {n}_i) x_i({k}) - \rho_z d_i {n}_i x_i({k-1})$ \\[1mm]
\hspace*{17.5mm} $+\ {n}_i \Big[ \rho_y  z_i(k) + \rho_z \sum_{j \in \mathcal{N}_{i}} \tilde{s}_{j}(k)  \Big]$ \\[1mm]
\hspace*{2.5mm} $\bar{x}_{i}({k+1}) = \sum_{\ell = 0}^{\Ccal-1}\alpha_{\ell}x_i(k+1-\ell), \ \sum_{\ell} \alpha_{\ell} = 1$ \\[1mm]
\hspace*{2.5mm} $y_{i}({k+1}) = \mathrm{max}(x_i({k+1})+\bar{u}_i(k),a_i)$ \\[1mm]
\hspace*{2.5mm} $\bar{u}_i({k+1}) = \bar{u}_i(k) + x_i({k+1}) - y_i({k+1} )$ \\[1mm]
\hspace*{2.5mm} $z_i(k+1) = 2 y_i({k+1}) - y_i({k})$ \\[1mm]
\hspace*{2.5mm} Send ${s}_{i}(k+1) = 2 \Bar{x}_{i}({k+1}) - x_i(k)$ to neighbors $j \in \mathcal{N}_{i}$ \\[1mm]
\textbf{EndFor}\\
\caption{The RD-MC algorithm.}\label{Alg1}
\end{algorithm} 

\section{Simulation Results} \label{Simul} 

\begin{figure}[t!]
    \centering
    \includegraphics[width=.45\textwidth]{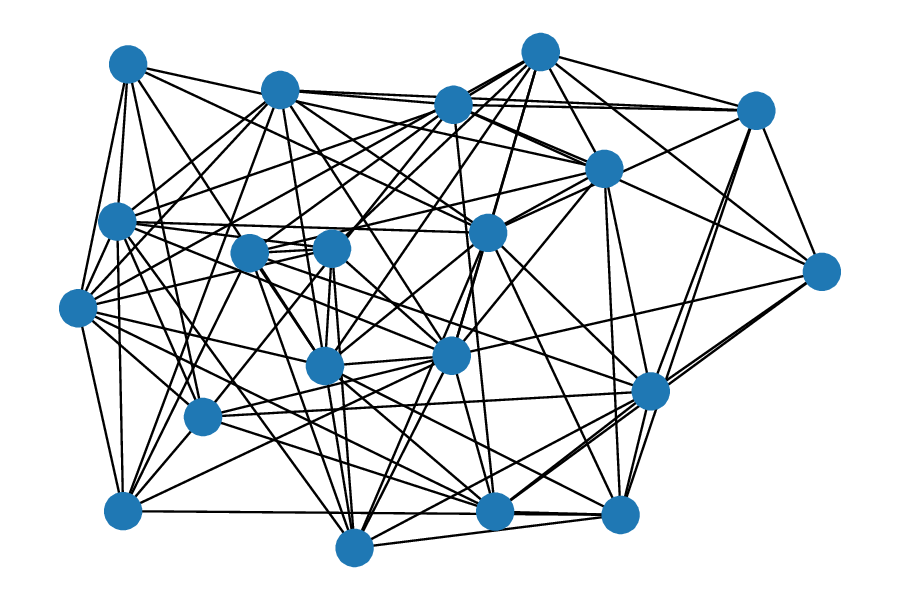}
    \caption{The considered network with an arbitrary topology and $J = 20$ agents.}
    \label{fig:fig1}
\end{figure}
\begin{figure}[t!]
    \centering
    \includegraphics[width=.45\textwidth]{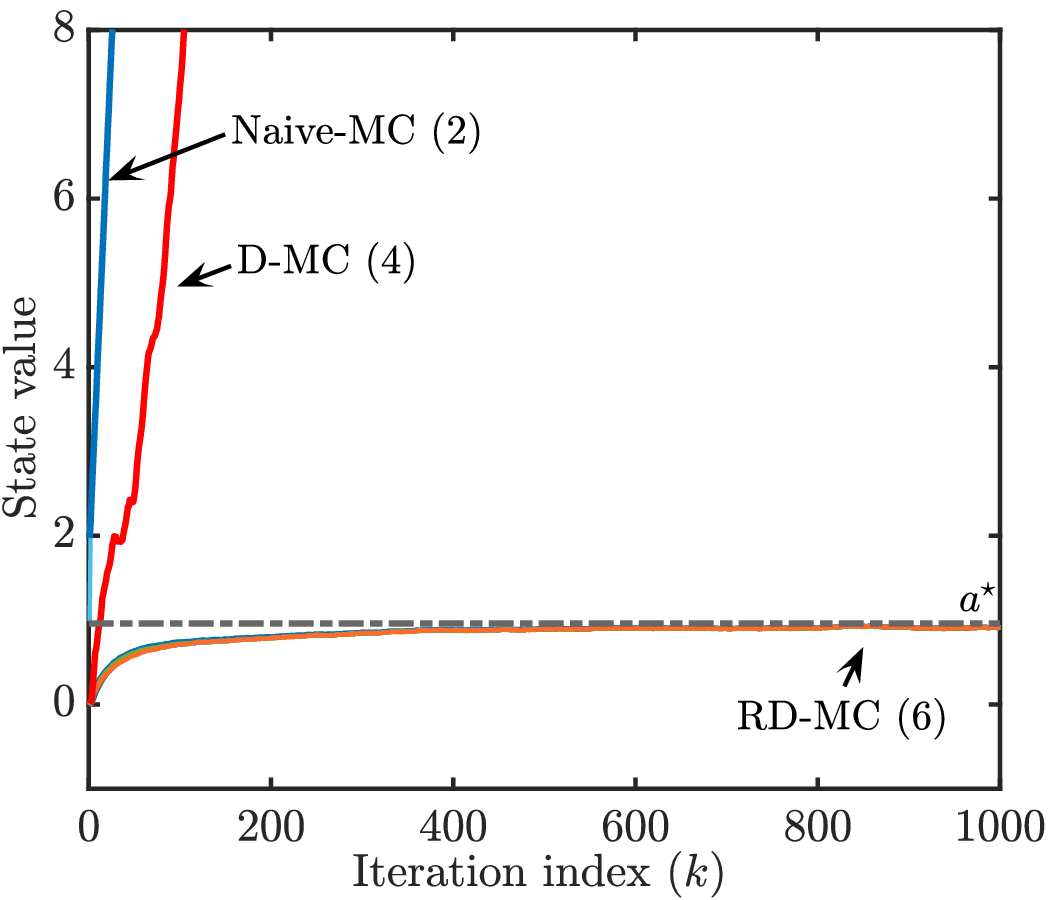}
    \caption{The impact of noise on the performance of naive-MC \eqref{eq:convmaxcons}, D-MC algorithm \eqref{eq:FDUp} and RD-MC algorithm \eqref{eq:FDRobust} with window size $\Ccal = 3$ and noise variance $\sigma^2 = 0.1$.}
    \label{fig:fig3}
\end{figure}
\begin{figure}[t!]
    \centering
    \includegraphics[width=.45\textwidth]{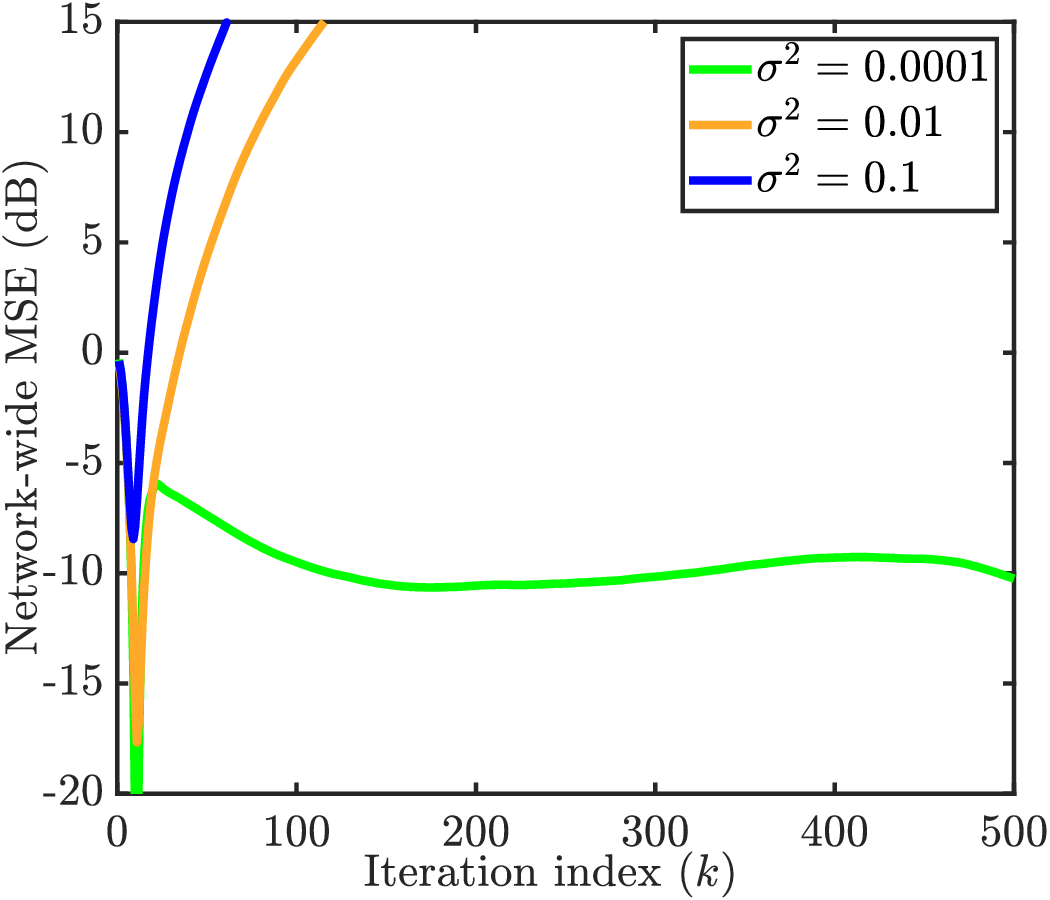}
    \caption{The effect of noise variance on the steady-state network-wide MSE of RD-MC with window size $\Ccal = 1$ and different noise variances $\sigma^2 \in \{ 0.0001, 0.01, 0.1\}$.}
    \label{fig:fig4c1}
\end{figure}
\begin{figure}[t!]
    \centering
    \includegraphics[width=.45\textwidth]{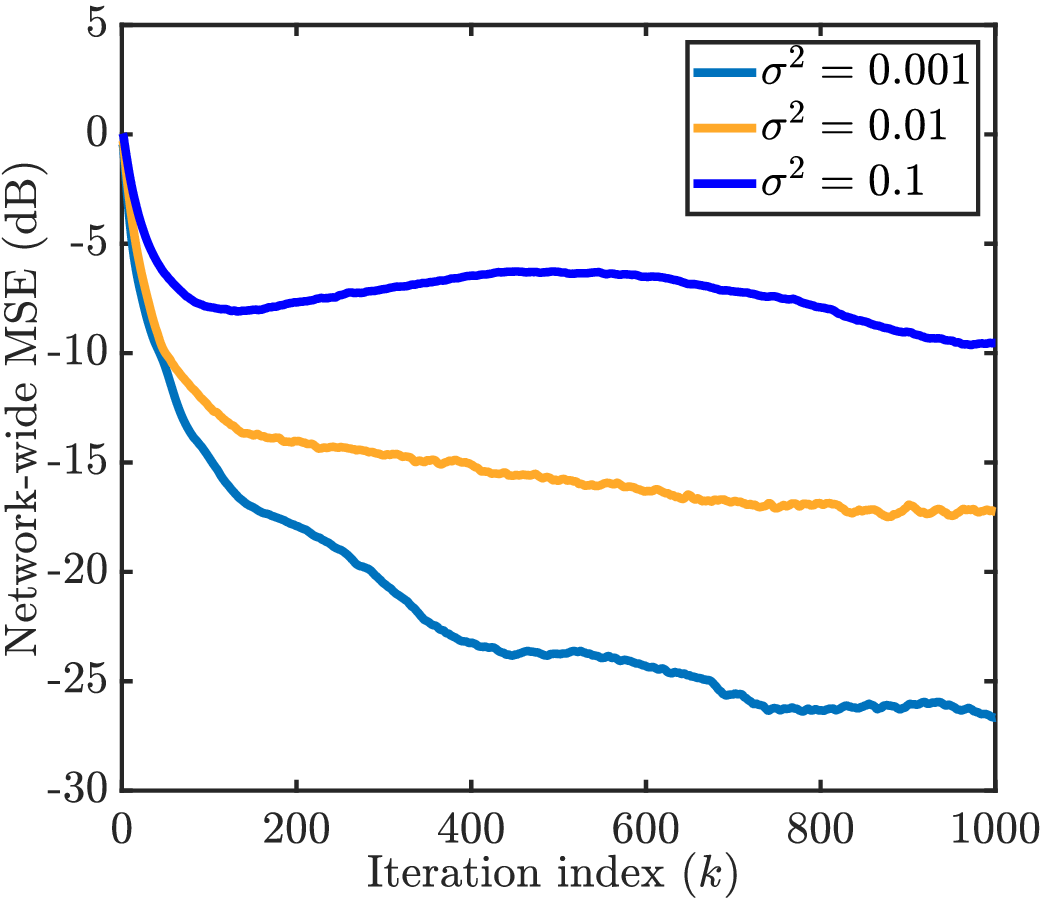}
    \caption{The effect of noise variance on the steady-state network-wide MSE of RD-MC with window size $\Ccal = 2$ and different noise variances $\sigma^2 \in \{ 0.001, 0.01, 0.1\}$.}
    \label{fig:fig4c2}
\end{figure}
\begin{figure}[t!]
    \centering
    \includegraphics[width=.45\textwidth]{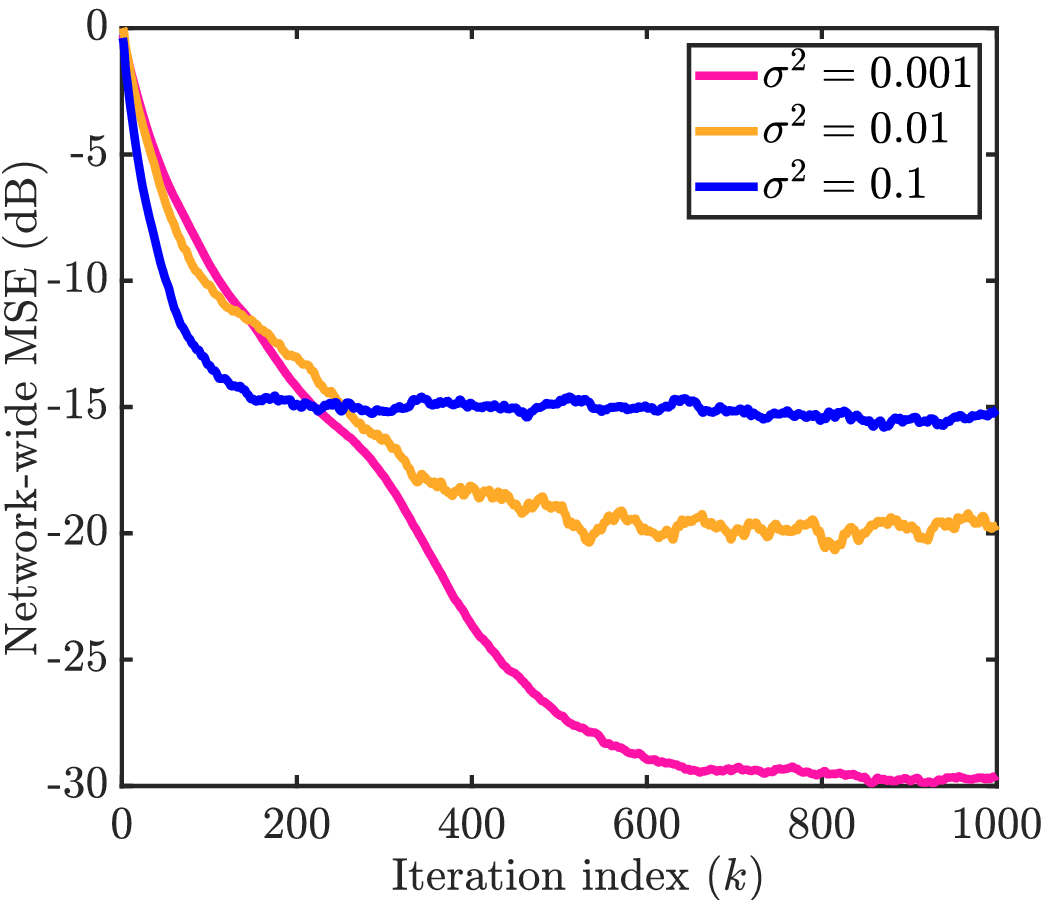}
    \caption{The effect of noise variance on the steady-state network-wide MSE of RD-MC with window size $\Ccal = 3$ and different noise variances $\sigma^2 \in \{ 0.001, 0.01, 0.1\}$.}
    \label{fig:fig4c3}
\end{figure}
\begin{figure}[t!]
    \centering
    \includegraphics[width=.45\textwidth]{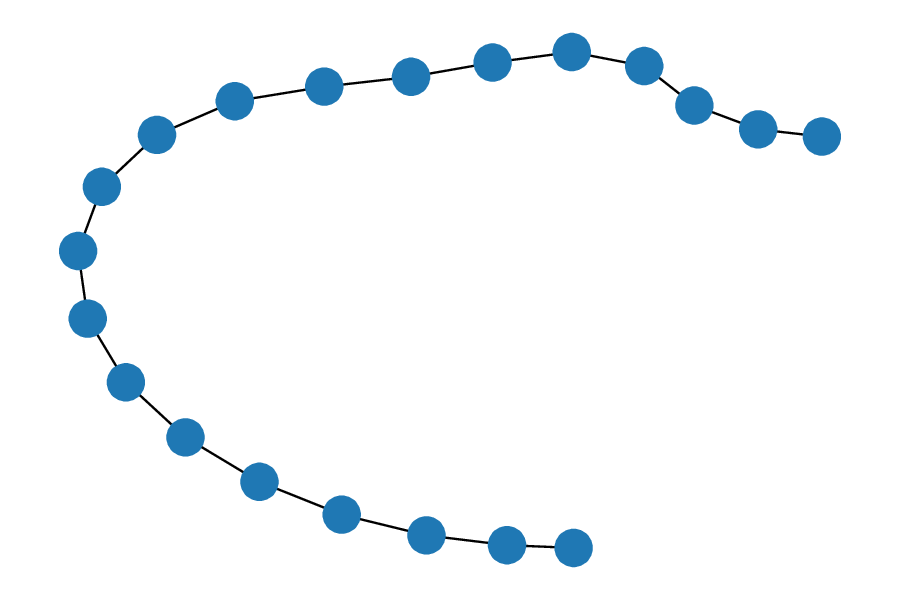}
    \caption{The considered network with linear topology and $J = 20$ agents.}
    \label{fig:linear}
\end{figure}
\begin{figure}[t!]
    \centering
    \includegraphics[width=.45\textwidth]{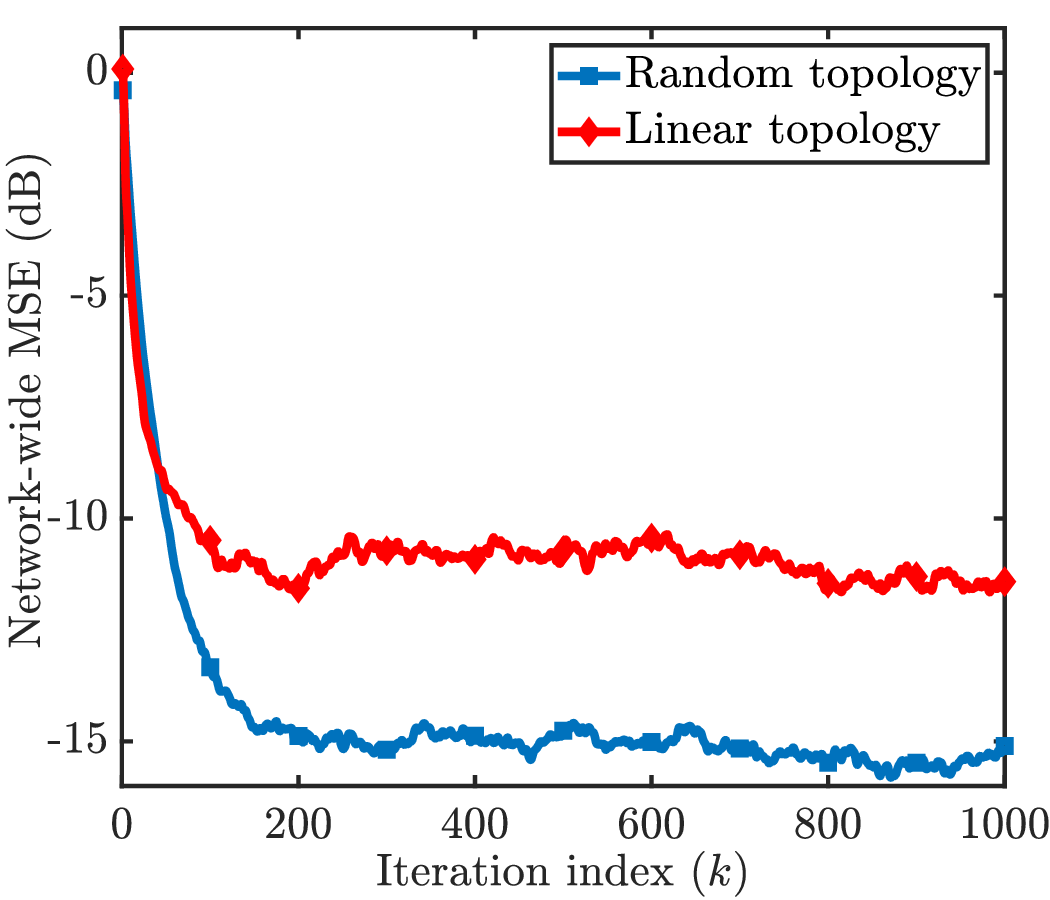}
    \caption{The impact of network connectivity on the network-wide MSE of RD-MC with window size $\Ccal = 3$ in the presence of link noise with variance $\sigma^2 = 0.1$.}
    \label{fig:fig6}
\end{figure}

We conduct a series of experiments to examine the performance of the proposed RD-MC algorithm. We consider a network of $J = 20$ agents as depicted in Fig. \ref{fig:fig1}. We independently draw the initial values (estimates) of the agents from a standard normal distribution, i.e., $a_i \sim \Ncal(0,1)\ \forall i\in\mathcal{V}$, and set $a^{\star}=\max(\{a_i\}_{i\in\mathcal{V}})$. In addition, we set the penalty parameter values to $\rho_z = \rho_y = 1$ and the weights in \eqref{eq:FDxBarUp} to $\alpha_{\ell} = 1/\Ccal$ in all our experiments. We obtain the results by averaging over $1000$ independent realizations of communication noise. To model the noise in the communication links, we employ a truncated zero-mean normal distribution, truncating the noise at $\pm 3 \sigma$ to ensure it remains within a reasonable bound.


In our first experiment, we examine the impact of noise on the performance of RD-MC, D-MC, and the naive solution \eqref{eq:convmaxcons}, referred to as naive-MC, using a noise variance of $\sigma^2 = 0.1$ and a window size of $\Ccal = 3$. Fig.~\ref{fig:fig3} shows the evolution of the estimates of all agents using the considered algorithms over $1000$ iterations. It is evident that RD-MC converges to the maximum value with a bounded error, whereas the other two algorithms diverge.

In our second experiment, we study the effect of noise variance on the network-wide mean square error (MSE) of RD-MC calculated as
$$ \frac{1}{J} \sum_{i=1}^{J} \mathbb{E}\left[({x}_{i}(k) - a^{\star})^2\right].$$
We conduct simulations of RD-MC using different window sizes $\Ccal \in \{1,2,3\}$ and noise variances $\sigma^2$, and present the results in Figs.~\ref{fig:fig4c1}-\ref{fig:fig4c3}. We observe that increasing $\sigma^2$ results in higher steady-state network-wide error across all experiments. However, the choice of window size profoundly influences RD-MC's efficacy in mitigating communication noise. While RD-MC struggles to converge with $\Ccal=1$, it maintains convergence with $\Ccal\ge2$ and increasing $\Ccal$ enhances its robustness to noise. Figs.~\ref{fig:fig3} and~\ref{fig:fig4c3} show that RD-MC with $\Ccal=3$ exhibits significantly greater resilience to link noise compared to D-MC, without imposing any additional computational or communication overhead.


In our final experiment, we assess the sensitivity of RD-MC's performance to network topology in the presence of link noise. We simulate RD-MC with a window size of $\Ccal = 3$ for two networks, namely, the network in \ref{fig:fig1} and a network with linear topology depicted in Fig. \ref{fig:linear}. We also set the noise variance to $\sigma^2 = 0.1$. We present the results in Fig. \ref{fig:fig6}. We observe that, with a linear network topology, 
the network-wide steady-state MSE of RD-MC is larger compared to a network with an arbitrary topology and higher average degree. However, RD-MC continues to perform well in the linear network topology with low connectivity.

\section{Conclusion} \label{Conc}

We developed a distributed algorithm, called noise-robust distributed maximum consensus (RD-MC), to tackle the challenge of identifying the maximum value within an ad-hoc multi-agent network utilizing noisy communication channels. Unlike existing algorithms designed for ad-hoc networks, RD-MC exhibits robustness against additive communication noise. Our extensive simulation results demonstrated the effectiveness of RD-MC in different scenarios.

\balance
\bibliographystyle{IEEEtran}
\bibliography{biblio} 

\end{document}